\documentclass{jpsj-suppl}
\usepackage{txfonts}

\title{Strangeness Photoproduction on Quasifree Neutrons}

\author{Dominik \textsc{Werthm\"uller}$^{1}$ on behalf of the A2 Collaboration}

\inst{$^{1}$School of Physics and Astronomy, University of Glasgow, Glasgow G12 8QQ, United Kingdom}

\email{dominik.werthmueller@glasgow.ac.uk}

\recdate{September 30, 2015}

\abst{Strangeness photoproduction in the elementary reactions $\gamma N\rightarrow KY$
offers the possibility to study nucleon resonances in the mass region above 1.7
GeV, where the number of states and their properties are still under debate.
These reactions could allow the first ``complete'' experiment in pseudoscalar 
meson photoproduction due to the easier access to recoil polarization
observables in experiments. In addition to measurements on proton targets, a
full isospin decomposition of the $I=1/2$ electromagnetic amplitudes requires 
also data obtained from quasifree neutron targets. This contribution shows
first preliminary results on feasibility studies for such measurements at
the A2 experiment at MAMI.}

\kword{strangeness, photoproduction, quasifree, neutron}

\begin{document}
\maketitle

\section{Introduction}
Current experimental activities in light baryon spectroscopy
at experiments such as A2 (MAMI),
CB and BGO-OD (ELSA), 
CLAS (JLab), and LEPS (SPring-8) concentrate 
on the extraction of polarization observables for meson photoproduction
reactions using polarized beams and/or polarized targets. In principle, a
minimum set of eight carefully chosen observables would allow the
determination of an unambiguous solution in terms of the involved 
amplitudes for pseudoscalar meson production 
(``complete'' experiment) \cite{Chiang_Tabakin}. 
Because of finite uncertainties of the experimental data,
it is favorable to have information about as many observables as possible to
remove ambiguities. Regarding beam and target polarization
observables, a large number of results are expected to be published 
for nonstrange meson photoproduction in the coming years. 
The measurement of recoil polarization observables for these reactions 
is more challenging since it requires a dedicated polarimeter in the 
experimental setup. 
On the other hand a ``complete'' measurement cannot just be any 
eight observables but it is necessary that at least one observable
is from beam-recoil or target-recoil experiments.
These can be extracted more easily in strangeness
photoproduction reactions $\gamma N\rightarrow KY$ because of
the self-analyzing weak decay of the hyperons $Y$.
Therefore, the ``complete'' experiment could be achieved earlier 
with respect to the nonstrange sector. 
Nevertheless, measurements involving
strangeness suffer from lower cross sections and additional 
complications related to the detection of kaons and hyperons.
Consequently, the experimental database lacks of precision and
coverage in terms of photon energy and kaon center-of-mass
polar angles. This is especially true at threshold, where 
the theoretical interpretation is more straightforward, 
and the situation should be improved in this regard.

Due to the electromagnetic formation in
photoproduction, a reliable isospin decomposition of the amplitudes
is only possible when experimental data for all six possible
reactions are available:
\begin{center}
\begin{tabular}{@{}lll@{}}
(1) $\gamma p\rightarrow K^+ \Lambda$ & 
(2) $\gamma p\rightarrow K^+ \Sigma^0$ & 
(3) $\gamma p\rightarrow K^0 \Sigma^+$ \\
(4) $\gamma n\rightarrow K^0 \Lambda$ &
(5) $\gamma n\rightarrow K^0 \Sigma^0$ &
(6) $\gamma n\rightarrow K^+ \Sigma^-$ \\
\end{tabular}
\end{center}
Particularly for the reactions (4)--(6) on the neutron, the
experimental database is still sparse
\cite{KPSM_06,KPSM_10,K0N_08,K0N_10,K0N_12}.
Concerning the study
of resonance contributions in the $s$-channel, reactions (4) and (5)
are of special interest because there are no $K^0$ background terms
in the $t$-channel. 
Therefore, even if resonance contributions are small close to 
threshold, a good knowledge of those reactions could help
to constrain $u-$ and $t-$channel background terms
in the theoretical models \cite{MART_11,MART_14}. 
In addition, experimental data very
close above the reaction threshold provide a precision test
of Chiral Perturbation Theory including the strangeness
degree of freedom \cite{CHPT_97,CHPT_07}.
Finally, measurements of strangeness 
photoproduction on bound nucleons may offer the possibility to study the 
$KN$ and $YN$ potentials via final state interaction (FSI) \cite{KYD_13}.
A good understanding of the latter is crucial for the
study of hypernuclei for example.

For this contribution, existing experimental data from
the A2 experiment at MAMI was analyzed to investigate the
feasibility of precision measurements of strangeness
photoproduction focussing on $\gamma n\rightarrow KY$
using deuterium as quasifree neutron target.

\section{Experimental Setup}
The A2 experiment is located at the MAMI continuous-wave electron 
accelerator facility in Mainz (Germany) \cite{MAMI_76,MAMI_08}.
A high intensity photon beam
is created from the 1.5 GeV primary beam by the bremsstrahlung 
tagging technique using the Glasgow photon tagger
\cite{TAGG_91,TAGG_96,TAGG_08}. After beam collimation
the photons impinge on the liquid deuterium target 
installed in the center of the sphere-like
electromagnetic calorimeter Crystal Ball (CB) \cite{CB_01}. This detector consists of
672 NaI(Tl) crystals and provides basic tracking via a multi-wire
proportional chamber (MWPC) and charged particle discrimination
via a $dE/E$ analysis
using a cylinder of plastic scintillator strips (PID \cite{PID}) surrounding the target.
The forward hole of CB is covered by the hexagon-shaped TAPS detector
wall \cite{TAPS_91,TAPS_94}. It comprises 366 BaF$_2$ and
72 PbWO$_4$ crystals.
Charged particles can be vetoed by plastic scintillator
tiles installed in front of the crystals. Particle identification
can be achieved via $dE/E$, time-of-flight, and pulse-shape
(slow/fast scintillation light component in BaF$_2$)
analyses.
The experimental trigger used 
to obtain the data for this work consisted of an energy
sum condition ($E_{\mathrm{tot}} > 300$ MeV in CB) and a condition ($>2$)
on the activated logical sectors (45 sectors in CB, 6 sectors in TAPS).

\section{Analysis}

\begin{table}[b]
\caption{Overview of the event selection for the analyzed 
$\gamma n\rightarrow KY$ reactions.}
\label{tabel:ev_sel}
\begin{tabular}{lll}
\hline\noalign{\smallskip}
Reaction & Particle Decays & Detected Particles \\ \noalign{\smallskip}
\hline\noalign{\smallskip}
(4)\, $\gamma n\rightarrow K^0\Lambda$  & 
$K^0_S\rightarrow\pi^0\pi^0$, $\Lambda\rightarrow p\pi^-$ & 
$4\gamma p\pi^-$ \\ \noalign{\smallskip}
(5)\, $\gamma n\rightarrow K^0\Sigma^0$ &  
$K^0_S\rightarrow\pi^0\pi^0$, $\Sigma^0\rightarrow \gamma\Lambda$ & 
$5\gamma p\pi^-$ \\ \noalign{\smallskip}
(6)\, $\gamma n\rightarrow K^+\Sigma^-$ & 
$K^+\rightarrow\mu^+\nu_\mu$ (in crystal), $\Sigma^-\rightarrow n\pi^-$ & 
$K^+ n\pi^-$ \\ \noalign{\smallskip}
\hline
\end{tabular}
\end{table}

\subsection{Event Selection}
An overview of the event selection for the three analyzed
$\gamma n\rightarrow KY$ reactions can be found
in table \ref{tabel:ev_sel}. The neutral $K^0$ was identified
via the $K^0_S\rightarrow\pi^0\pi^0$ decay 
($\Gamma_i/\Gamma = 30.69\%$ \cite{PDG}). Therefore, four
photons from the $K^0_S$ decay were requested for 
reactions (4) and (5). An additional photon was requested in
the selection for reaction (5) coming from the radiative
decay $\Sigma^0\rightarrow\gamma\Lambda$ of the $\Sigma^0$ 
hyperon ($\Gamma_i/\Gamma \approx 100\%$ \cite{PDG}).
The $\Lambda$ hyperon present in both reactions was detected
via demanding a proton and a pion originating from 
the $\Lambda\rightarrow p\pi^-$ decay 
($\Gamma_i/\Gamma = 63.9\%$ \cite{PDG}). Charged and neutral
particles were discriminated via the PID and veto detectors.
Proton and pion candidates were identified via a $dE/E$ analysis
(no charge separation possible for pions).

Event candidates for reaction (6) were filtered via
the presence of an in-crystal muonic decay of a $K^+$ meson
(see next subsection) plus
a coincident neutral cluster (neutron candidate), and
a charged pion cluster identified via a $dE/E$
analysis ($\pi^-$ candidate).

\begin{figure}[t]
\centering
\includegraphics[width=0.32\textwidth]{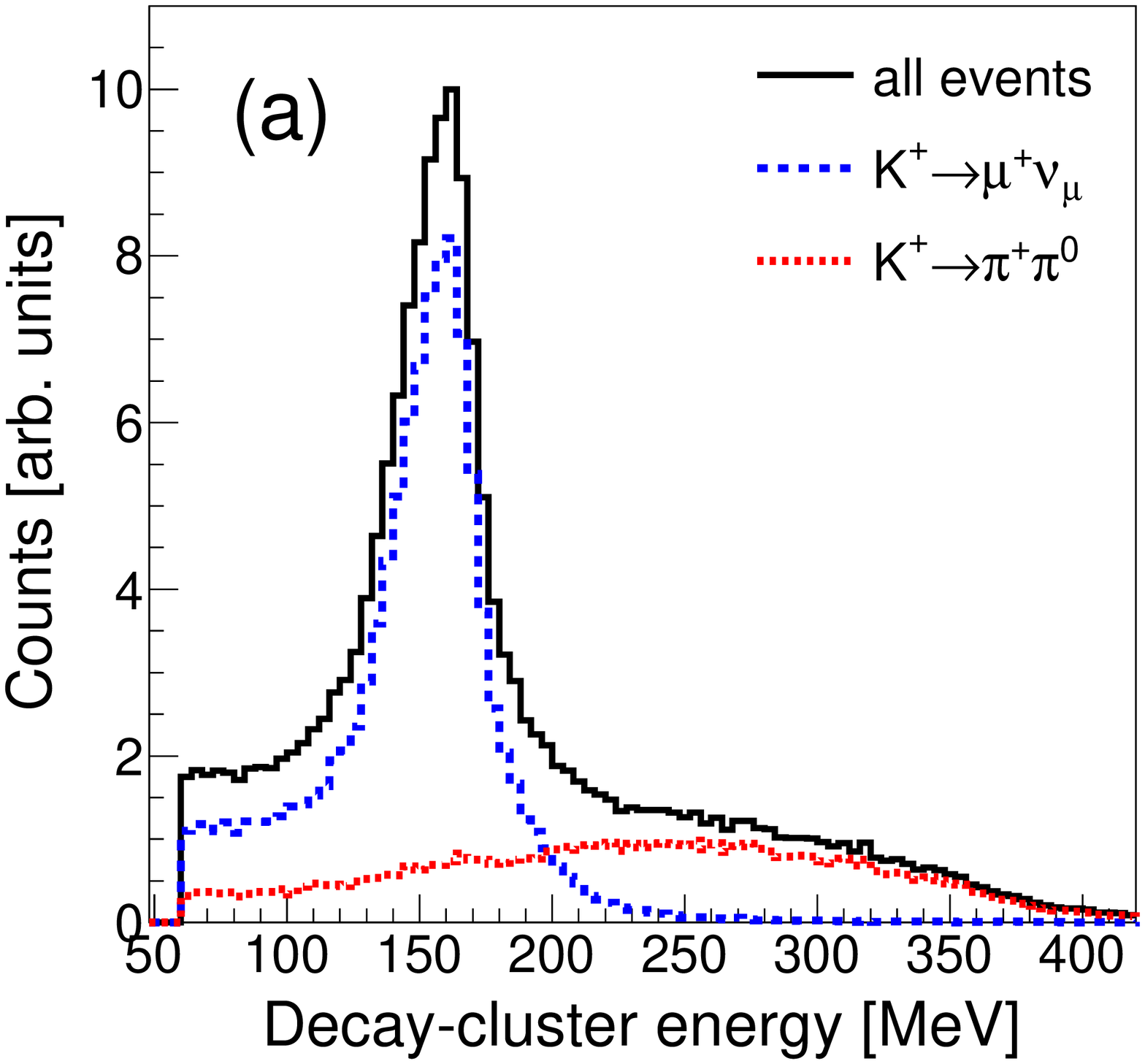}
\includegraphics[width=0.32\textwidth]{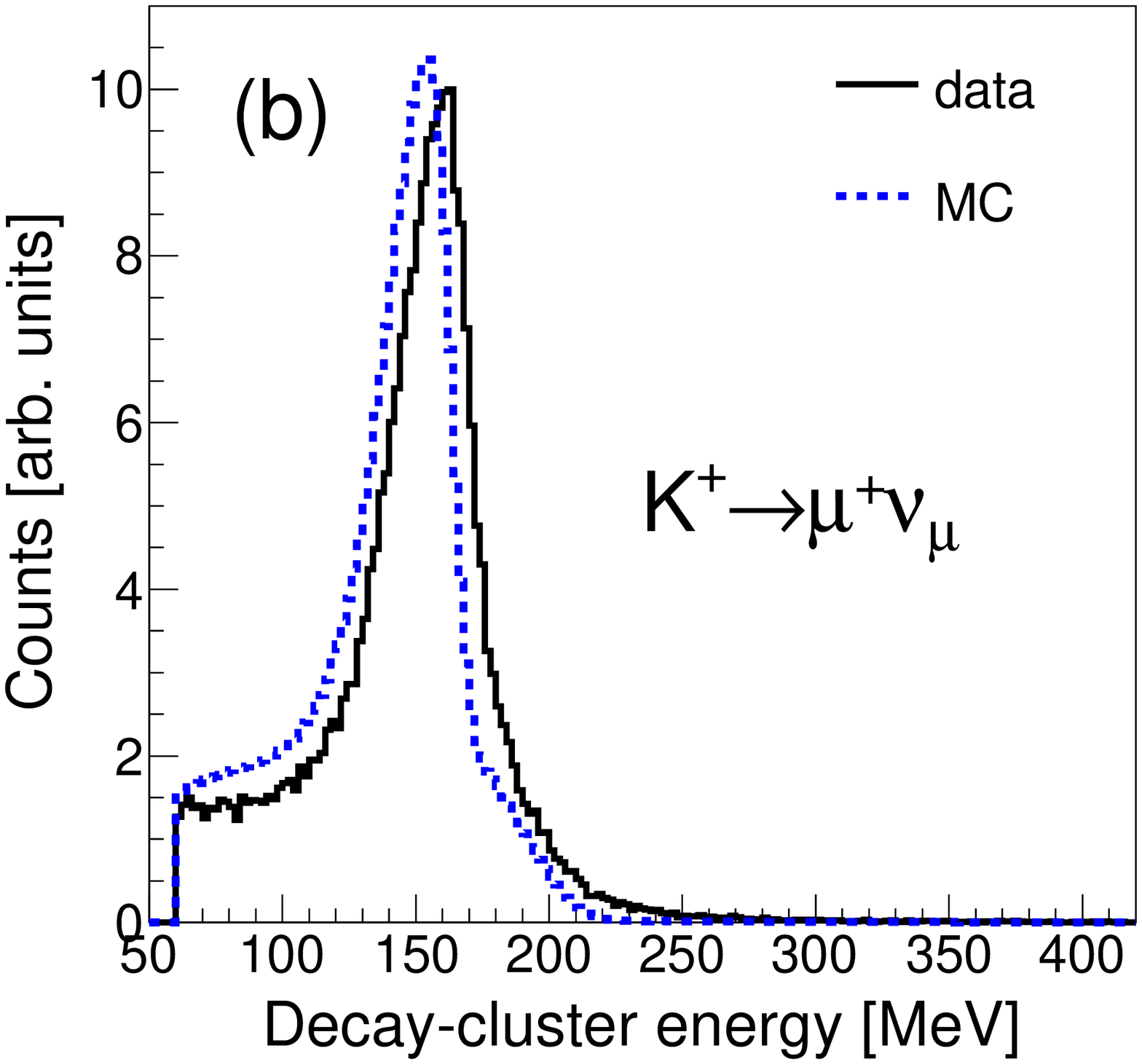}
\includegraphics[width=0.32\textwidth]{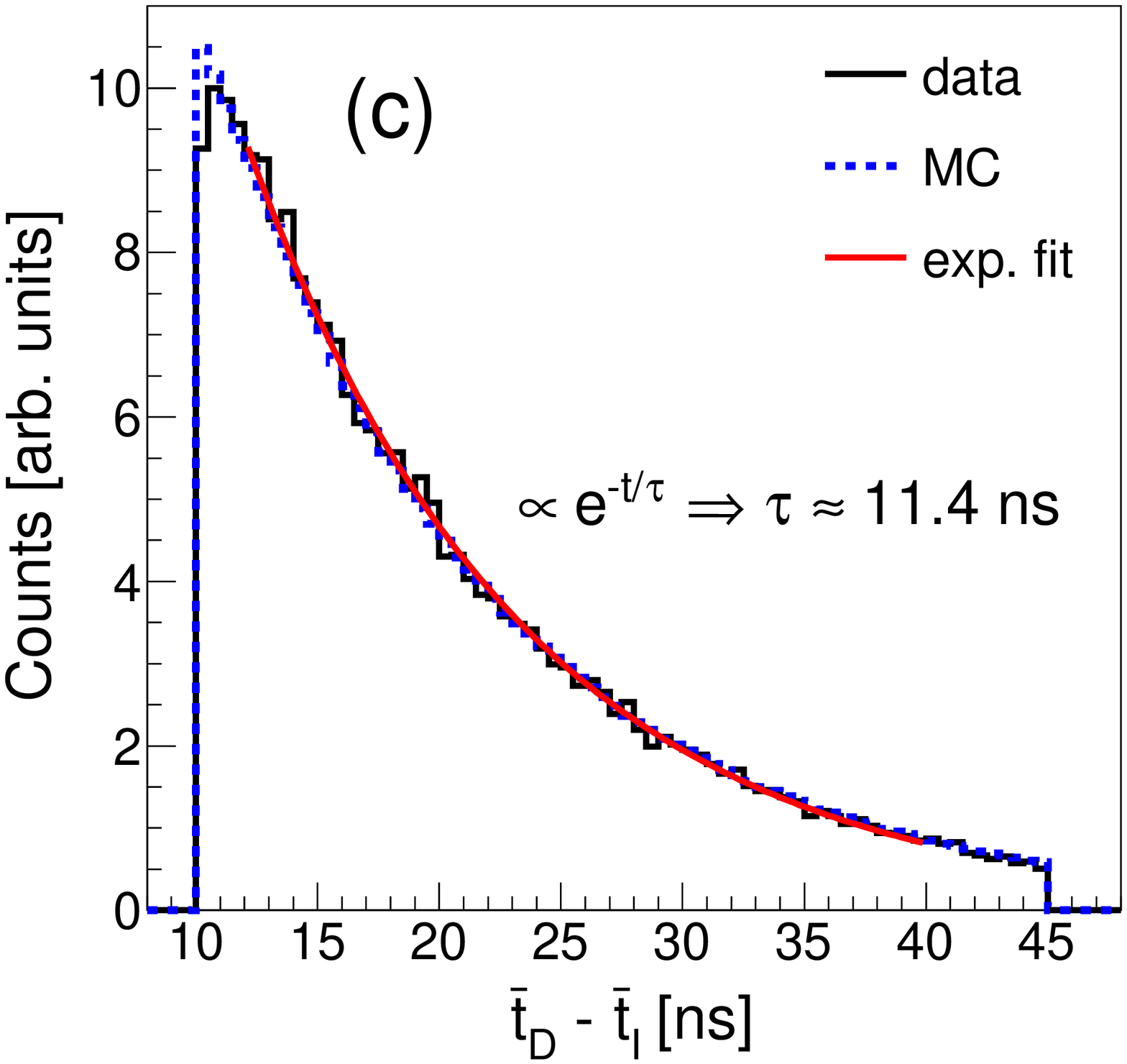}
\caption{$K^+$ detection in the Crystal Ball:
(a) Energy distribution of the $K^+$ decay clusters 
of all events (black solid), pionic decay
events (blue dotted) and muonic decay events
(red dashed).
(b) Comparison of experimental (black solid) and simulated
(blue dotted) decay cluster energy distributions for muonic decay
events.
(c) Impact-decay cluster time difference:
Experimental (black solid) and simulated (blue dotted)
distributions. Exponential fit to experimental
data (red curve).}
\label{fig:kaon_id}
\end{figure}

\subsection{$K^+$ Detection via In-Crystal Decay}
$K^+$ mesons were detected using their decay inside the
NaI(Tl) crystal of CB after a mean lifetime of 12.38 ns
\cite{PDG} using the technique described
in \cite{Jude_14}. This technique splits potential $K^+$
clusters into impact and decay subclusters based on 
the timing signals. Energy
and direction of the kaon can then be accessed
via the impact cluster, while properties of the
decay cluster help to differentiate between
the dominant
$\mu^+\nu_\mu$ ($\Gamma_i/\Gamma = 63.55\%$) and
$\pi^+\pi^0$ ($\Gamma_i/\Gamma = 20.66\%$) \cite{PDG}
decays.

Fig.~\ref{fig:kaon_id} shows some characteristic
distributions related to the $K^+$ detection technique.
In (a), the decompositon of the total decay cluster
energy distribution (black histogram)
into the contributions
of the muonic (blue histogram) and the pionic 
(red histogram) decays is illustrated. A clear peak around
150 MeV due to the energy deposited by the $\mu^+$
can be seen, whereas the distribution coming from 
the pionic decay is broader. Fig.~\ref{fig:kaon_id}(b)
shows the good agreement of the experimental
and the simulated distributions in case of the muonic decay.
Only an overall energy loss correction seems to be
necessary to account for the systematic shift.
Finally, the time difference between the impact
and the decay cluster is plotted in Fig.~\ref{fig:kaon_id}(c)
for experimental and simulated data. As expected, both 
distributions follow an exponential drop-off and
the decay time extracted from the experimental data
is in good agreement with the mean lifetime of the $K^+$ meson.

\subsection{Particle Reconstruction and Analysis Cuts}
The $\Lambda$ hyperon was reconstructed from the
detected proton and $\pi^-$ candidates. Since the
energy calibration of the calorimeters was optimized
for photons, and the relation between kinetic energy
and deposited energy is nonlinear for low energetic 
hadrons, an energy correction needs to be 
determined and applied. In addition, 
a cut-off in terms of deposited energy has to be
added to remove punch-through
particles, which deposit less than their kinetic energy.
For $\pi^-$ mesons, an additional cut rejecting 
clusters consisting of more than four crystals was 
applied to improve the energy resolution by removing
energy depositions from secondary reactions \cite{Marin_98}.

\begin{figure}[t]
\centering
\includegraphics[width=0.32\textwidth]{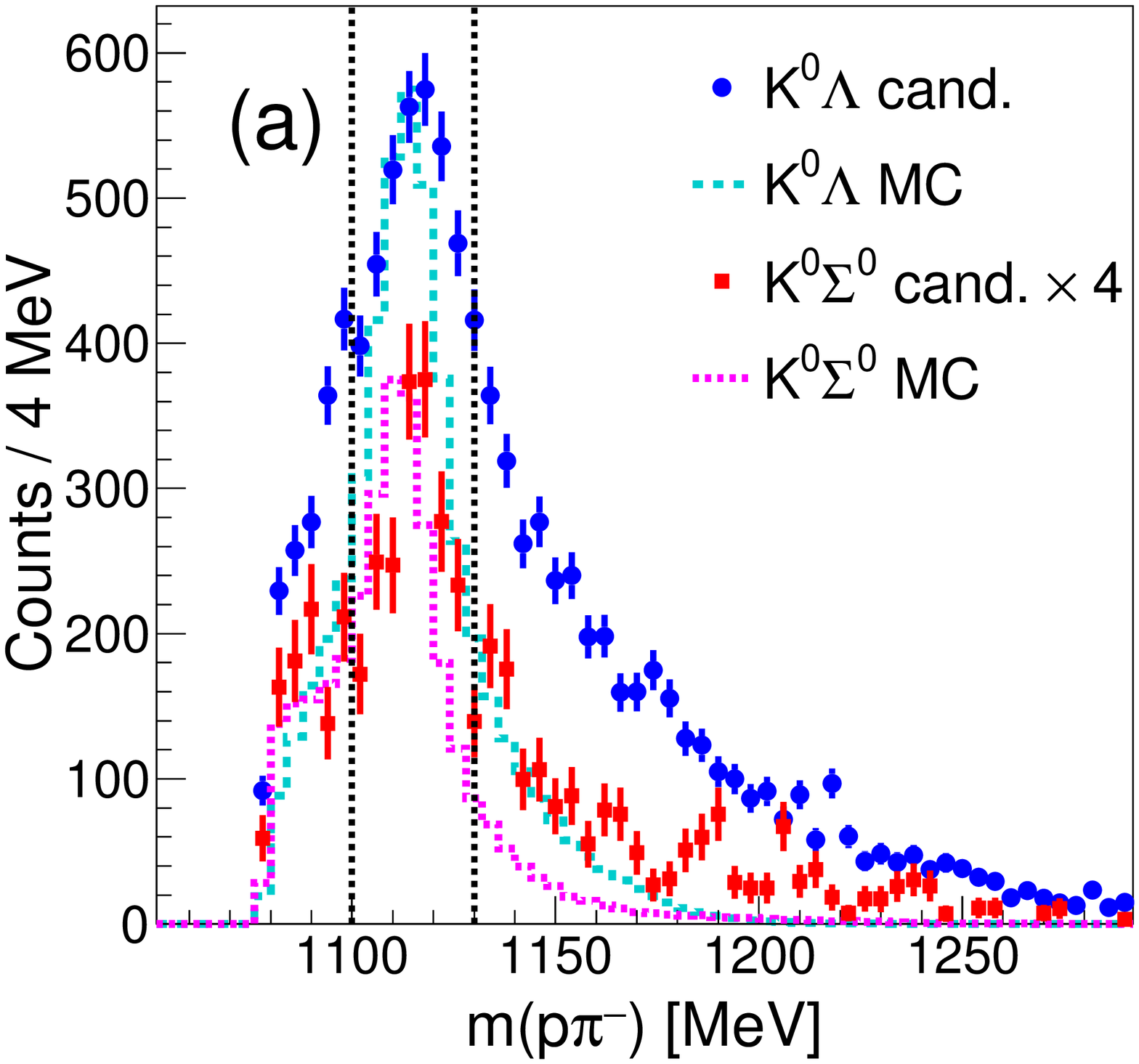}
\includegraphics[width=0.32\textwidth]{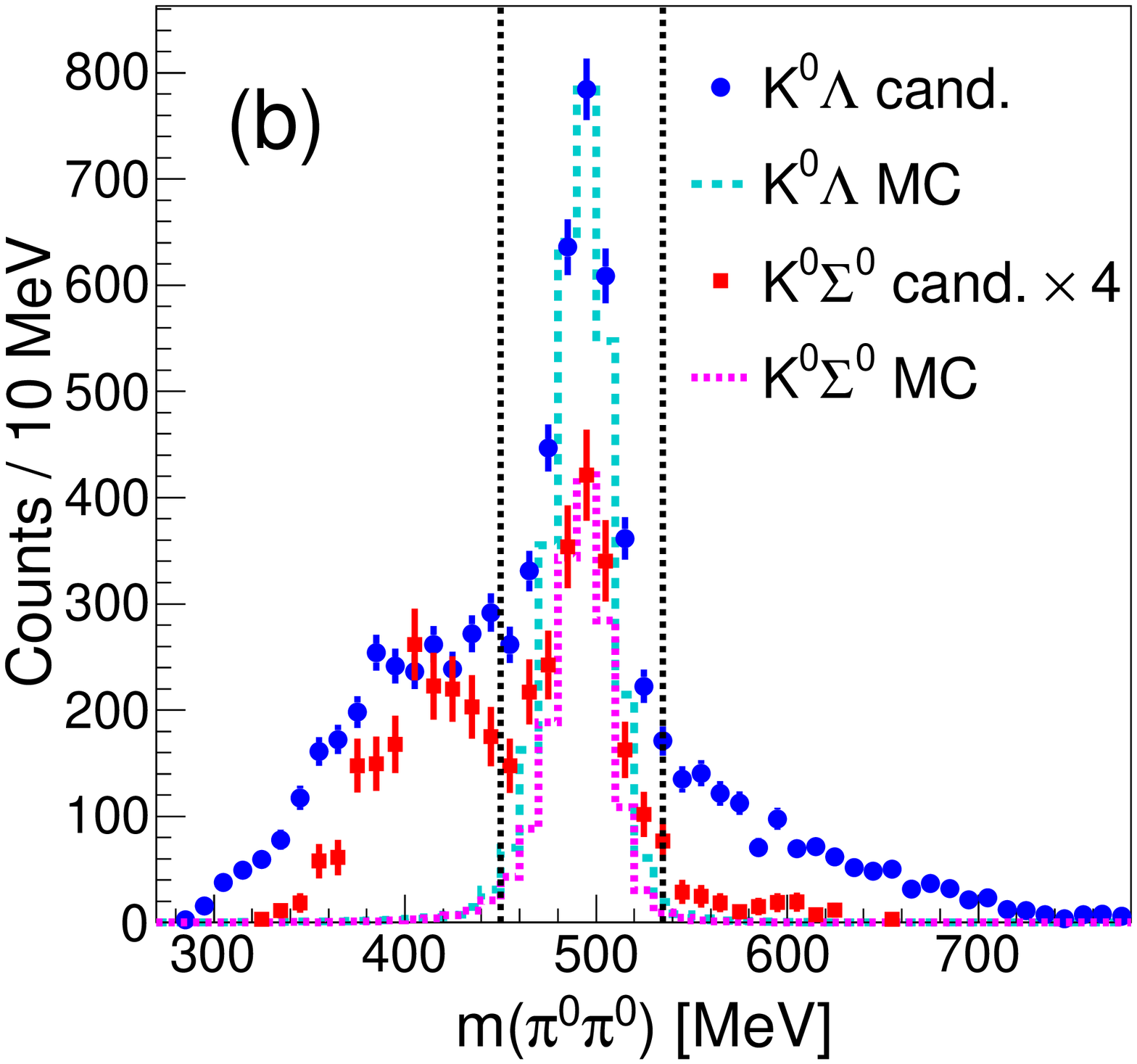}
\includegraphics[width=0.32\textwidth]{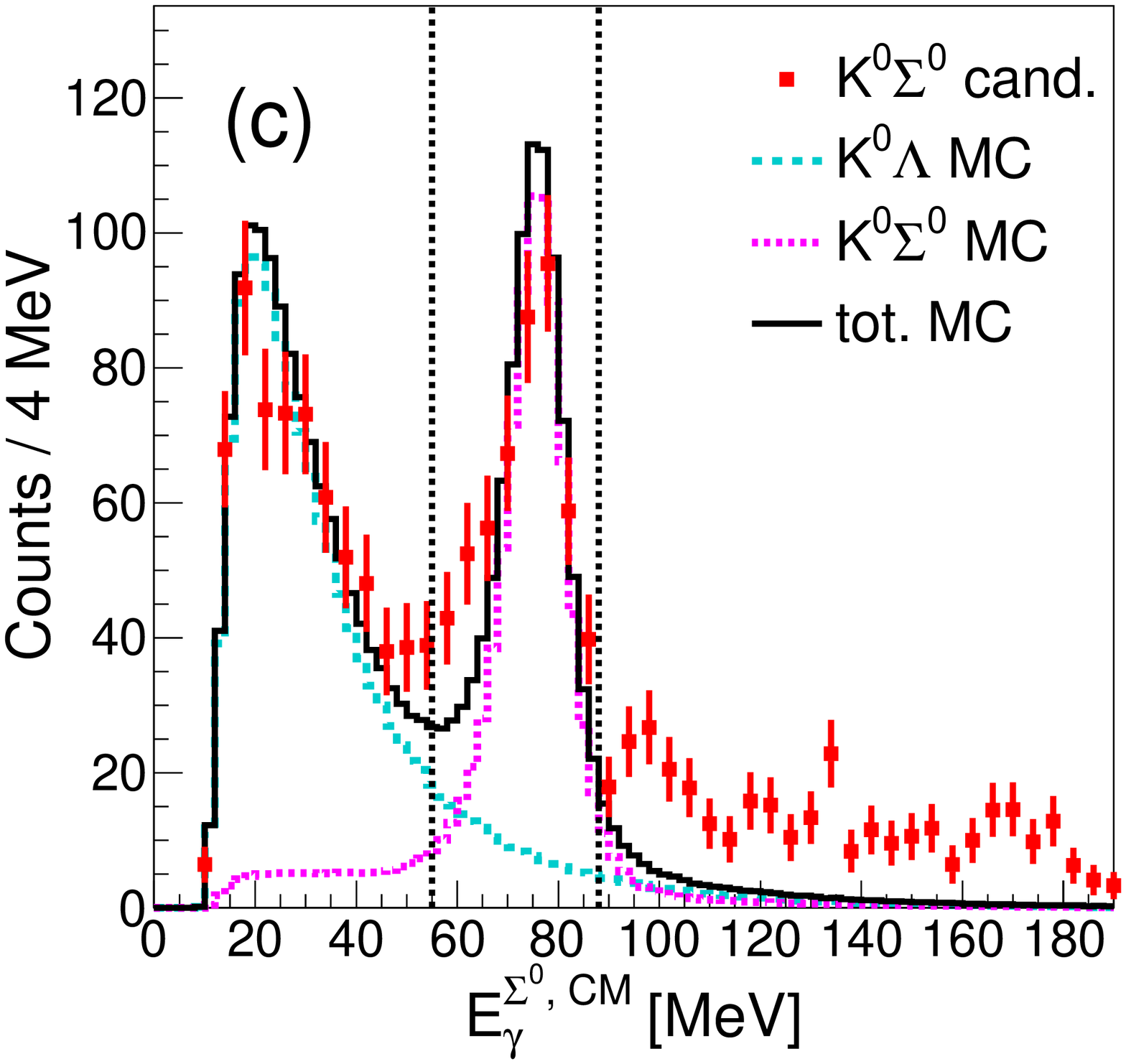}
\caption{Particle reconstruction distributions: 
Color code: blue circles (red squares): experimental
distributions for $K^0\Lambda$ ($K^0\Sigma^0$) event
candidates. Cyan dashed (magenta dotted) histograms:
simulated distributions for $K^0\Lambda$ ($K^0\Sigma^0$)
events.
(a) $p\pi^-$ invariant mass showing the $\Lambda$ signal.
(b) $\pi^0\pi^0$ invariant mass showing the $K^0$ signal.
(c) Energy distribution of the decay photon from 
$\Sigma^0\rightarrow\gamma\Lambda$
in the rest frame of the $\Sigma^0$ hyperon.
}
\label{fig:reco}
\end{figure}

A clear peak around the nominal
$\Lambda$ mass of about 1116 MeV can be seen
in the $p\pi^-$ invariant mass of Fig.~\ref{fig:reco}(a).
The agreement
between the experimental and simulated distribution
seems to be quite good for the more selective $K^0\Sigma^0$
event candidates, whereas some more background is remaining
in the $K^0\Lambda$ events. All complementary analysis
cuts were applied to obtain the presented distributions.
A cut around the $\Lambda$ peak (vertical lines) was
used to select good event candidates.

The neutral kaon was reconstructed from the four 
detected decay photon candidates using a kinematic fit
testing the hypothesis 
$K^0\rightarrow\pi^0\pi^0\rightarrow4\gamma$. All
photon combinations were fitted and the one with the
smallest $\chi^2$ was selected as good $K^0$ candidate.
As can be seen in Fig.~\ref{fig:reco}(b), a clear
peak coming from the $K^0$ is visible in the $\pi^0\pi^0$ invariant
mass distributions. The composition of the remaining
background events still needs to be studied.
All complementary analysis
cuts were applied to obtain the presented distributions.
A cut around the $K^0$ peak (vertical lines) and
additionally a cut in the $2\gamma$ invariant masses
selecting two good $\pi^0$ candidates (not shown here)
were introduced in the analysis.

The additional photon detected for the $K^0\Sigma^0$
event candidates was checked to originate from the
$\Sigma^0\rightarrow\gamma\Lambda$ decay by calculating
its energy in the rest frame of the $\Sigma^0$ candidate.
The resulting distribution shown in Fig.~\ref{fig:reco}(c)
reveals a clear peak around 76 MeV consistent with
the $\Sigma^0$-$\Lambda$ mass difference, and agrees with
the distribution obtained from simulation. The background
contribution at low energies is nicely described by the
$K^0\Lambda$ simulation where probably artificial photon
clusters from split-off effects are generated.
All complementary analysis
cuts were applied to obtain the presented distributions.
A cut around the peak (vertical lines) was
used to select good event candidates.

\begin{figure}[thb]
\centering
\includegraphics[width=\textwidth]{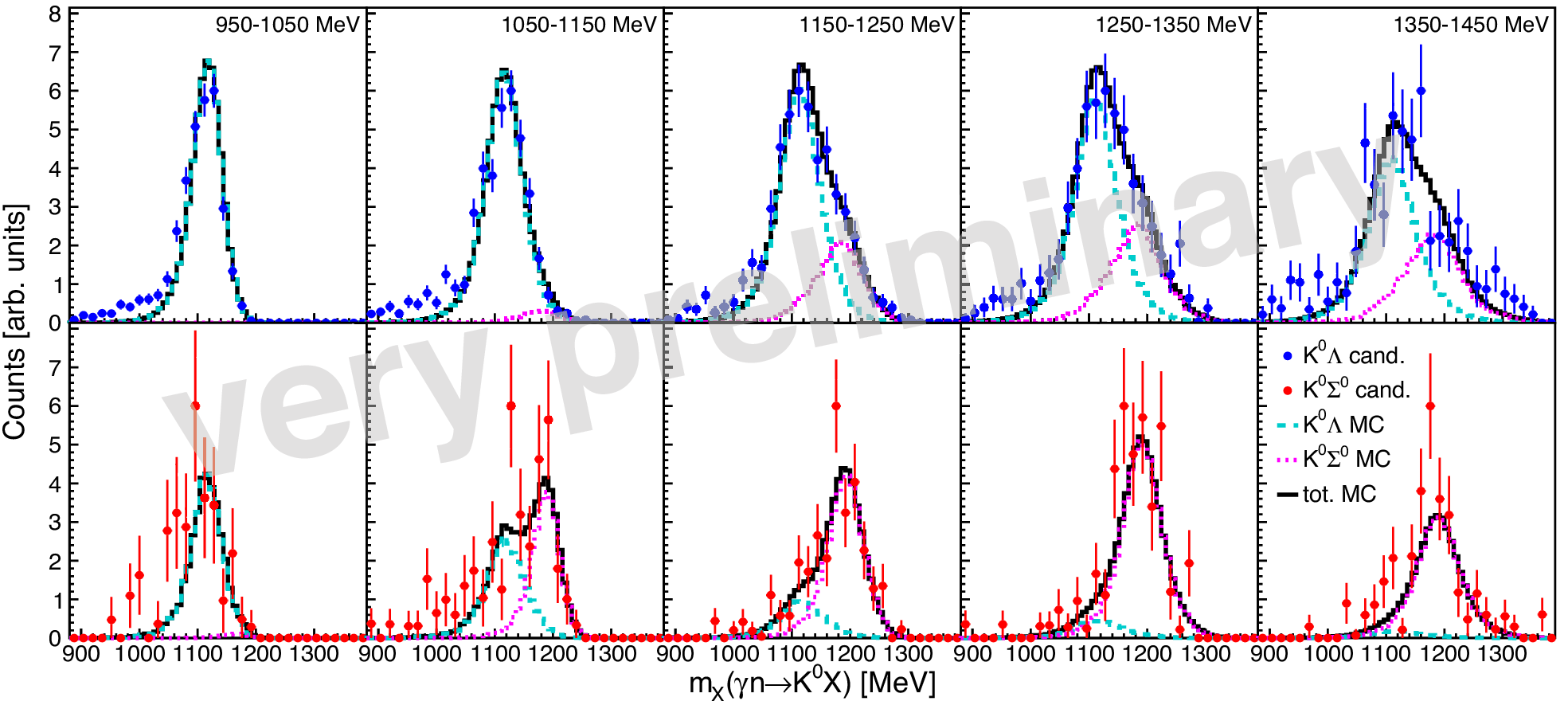}
\caption{Missing mass distributions of $\gamma n\rightarrow K^0X$
for different bins of incoming photon beam energy:
Color code: blue circles (red squares): experimental
distributions for $K^0\Lambda$ ($K^0\Sigma^0$) event
candidates. Cyan dashed (magenta dotted) histograms:
simulated distributions for $K^0\Lambda$ ($K^0\Sigma^0$)
events.
Upper row: $K^0\Lambda$ event candidates. 
Lower row: $K^0\Sigma^0$ event candidates.
}
\label{fig:k0_mm}
\end{figure}

The detection of the $\pi^-$ and
the neutron decay products of
the $\Sigma^-$ hyperon are rather 
challenging with the detectors of the A2
experiment. Neutron detection efficiencies in CB
are below 40\% \cite{NEFF_15} and the neutron kinetic 
energy can only be measured via time-of-flight in the
forward TAPS wall. Nevertheless, in case of the 
quasifree reaction on the neutron in 
$\gamma d\rightarrow K^+\Sigma^-(p)$,
by measuring the neutron direction along
with the directions and energies of the $K^+$ and
the $\pi^-$, it is possible to calculate the
neutron energy from kinematics \cite{Werthm_14},
although the resolution is additionally
smeared due to the secondary $\Sigma^-$ decay vertex.

Further cuts applied in the analysis were the 
condition for the kaon-hyperon coplanarity 
(requesting a difference of 180$^\circ$ in
the azimuthal angles in the lab frame) and 
the request for a missing proton in the 
$\gamma d\rightarrow KYX$ missing mass.

\section{Very Preliminary Results}
Very preliminary results in terms of missing mass
distributions for the $\gamma n\rightarrow K^0Y$ 
reactions are shown in Fig.~\ref{fig:k0_mm}.
The experimentally obtained distributions can
for most parts be reasonably described by
the sum of the simulated distributions 
of both reactions. Because the two event classes
differ only in the detection of the additional
decay photon of the $\Sigma^0$ hyperon, there
is contamination of the $K^0\Lambda$ and $K^0\Sigma^0$
final states in both directions. Therefore, the most
suited approach is probably a simultaneous yield
extraction for both reactions from the distributions
of the two event classes.

\begin{figure}[thb]
\centering
\includegraphics[width=\textwidth]{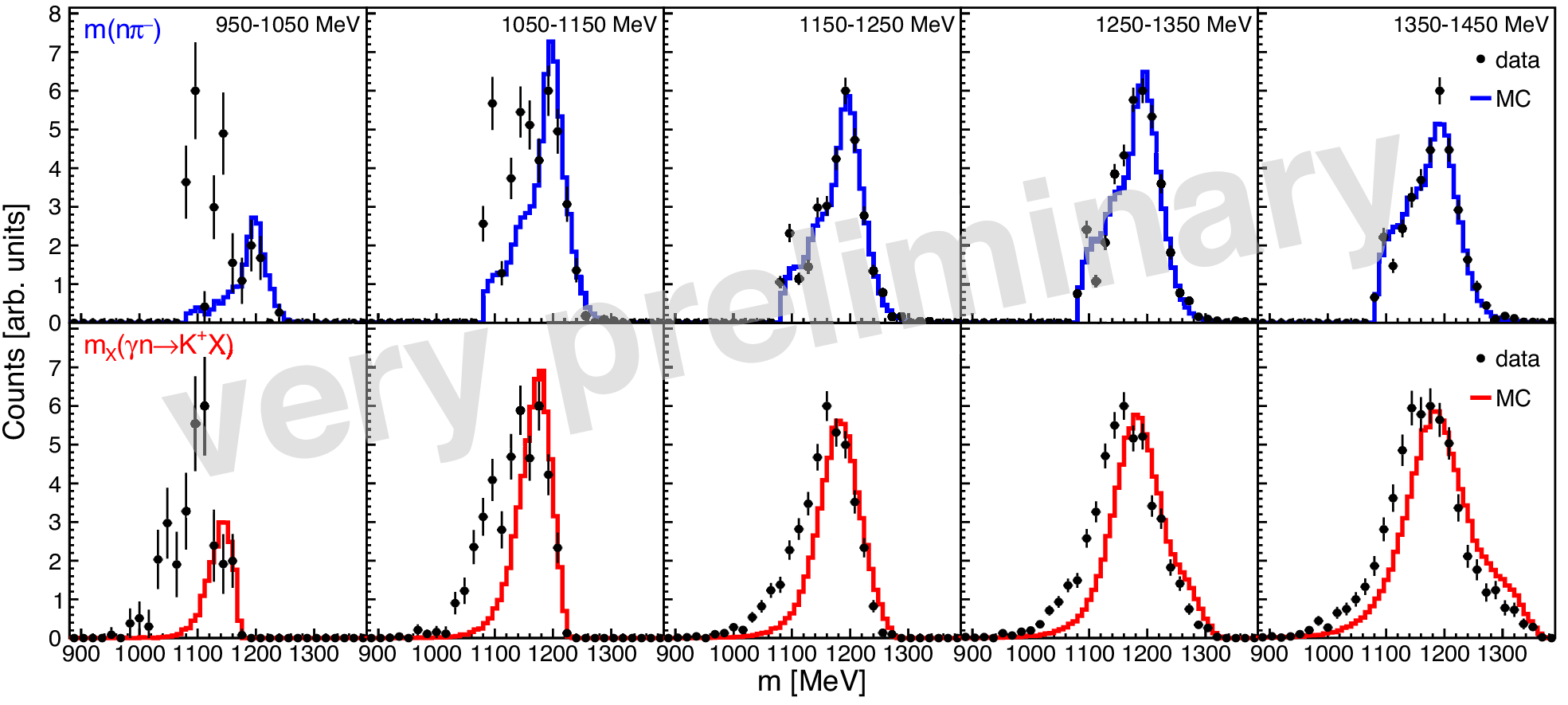}
\caption{$\Sigma^-$ mass distributions for different
bins of incoming photon beam energy:
Upper row: $n\pi^-$ invariant mass experimental
(black circles) and simulated (blue histograms) distributions.
Lower row: $\gamma n\rightarrow K^+X$ missing mass
experimental (black circles) and simulated 
(red histograms) distributions.
}
\label{fig:kplus_mm}
\end{figure}

In the analysis of the $\gamma n\rightarrow K^+ \Sigma^-$
reaction, the signal of the $\Sigma^-$ hyperon can be
searched in either the $n\pi^-$ invariant mass or the
$\gamma n\rightarrow K^+X$ missing mass. Both distributions
are shown in Fig.~\ref{fig:kplus_mm}. The resolution seems
to be comparable for lower photo beam energies, while
the resolution in the $n\pi^-$ invariant mass seems to be
better for higher energies. Besides a systematic
shift in the missing mass distributions due to the yet 
missing $K^+$ energy correction, the agreement between
the experimental and simulated distributions is 
surprisingly good considering the early 
stage of the analysis. The remaining
background in the lower photon energy bins still
needs to be investigated.

\section{Summary and Outlook}
Very preliminary analyses of all three isospin reactions 
$\gamma n\rightarrow KY$ in quasifree kinematics
using data from the A2 experiment obtained with a 
deuterium target give promising results regarding their identification.
More refinements and corrections,
especially concerning the detection of charged particles,
should further improve the results. After optimizing
the analyses in terms of statistical quality and systematic
uncertainties, it remains to be seen if cross section
and recoil polarization observables can be extracted
from the existing data or if new measurements should be
carried out.

\section{Acknowledgments}
This contribution and the participation at the NSTAR2015 conference was supported 
by the Swiss National Science Foundation (158822).


\begin{thebibliography}{9}
\bibitem{Chiang_Tabakin} W.-T. Chiang, F. Tabakin, Phys. Rev. C                                {\bf 55}  (1997) 2054.
\bibitem{KPSM_06} H. Kohri            {\it et al.}, Phys. Rev. Lett.                           {\bf 97}  (2006) 082003.
\bibitem{KPSM_10} S. Anefalos Pereira {\it et al.}, Phys. Lett. B                              {\bf 688} (2010) 289.
\bibitem{K0N_08} K. Tsukada           {\it et al.}, Phys. Rev. C {\bf 78} (2008) 014001; Phys. Rev. C {\bf 83} (2011) 039904.
\bibitem{K0N_10} H. Kanda             {\it et al.}, Nucl. Phys. A                              {\bf 835} (2010) 317.
\bibitem{K0N_12} Futatsukawa          {\it et al.}, EPJ Web of Conferences                     {\bf 20}  (2012) 02005.
\bibitem{MART_11} T. Mart,                          Phys. Rev. C                               {\bf 83}  (2011) 048203.
\bibitem{MART_14} T. Mart,                          Phys. Rev. C                               {\bf 90}  (2014) 065202.
\bibitem{CHPT_97} S. Steininger, U.G. Mei{\ss}ner,  Phys. Lett. B                              {\bf 391} (1997) 446.
\bibitem{CHPT_07} B. Borasoy          {\it et al.}, Eur. Phys. J. A                            {\bf 34}  (2007) 161.
\bibitem{KYD_13} P. Vancraeyveld      {\it et al.}, Nucl. Phys. A                              {\bf 897} (2013) 42.
\bibitem{MAMI_76} H. Herminghaus      {\it et al.}, Nucl. Instrum. Methods                     {\bf 138} (1976) 1.
\bibitem{MAMI_08} K.-H. Kaiser        {\it et al.}, Nucl. Instrum. Methods Phys. Res., Sect. A {\bf 593} (2008) 159.
\bibitem{TAGG_91} I. Anthony          {\it et al.}, Nucl. Instrum. Methods Phys. Res., Sect. A {\bf 301} (1991) 230.
\bibitem{TAGG_96} S.J. Hall           {\it et al.}, Nucl. Instrum. Methods Phys. Res., Sect. A {\bf 368} (1996) 698.
\bibitem{TAGG_08} J.C. McGeorge       {\it et al.}, Eur. Phys. J. A                            {\bf 37}  (2008) 129.
\bibitem{CB_01} A. Starostin          {\it et al.}, Phys. Rev. C                               {\bf 64}  (2001) 055205.
\bibitem{PID} D. Watts, Calorimetry in Particle Physics, Proceedings of the 11th International Conference, Perugia, Italy, 2004 (World Scientific, Singapore, 2005), 560.
\bibitem{TAPS_91} R. Novotny,                      IEEE Trans. Nucl. Sci.                      {\bf 38}  (1991) 379.
\bibitem{TAPS_94} A.R. Gabler         {\it et al.}, Nucl. Instrum. Methods Phys. Res., Sect. A {\bf 346} (1994) 168.
\bibitem{PDG} K.A. Olive              {\it et al.} (Particle Data Group), Chin. Phys. C        {\bf 38}  (2014) 090001.
\bibitem{Jude_14} T.C. Jude           {\it et al.}, Phys. Lett. B                              {\bf 735} (2014) 112.
\bibitem{Marin_98} A. Mar\'in         {\it et al.}, Nucl. Instrum. Methods Phys. Res., Sect. A {\bf 417} (1998) 137.
\bibitem{NEFF_15} M. Martemianov      {\it et al.}, JINST                                      {\bf 10}  (2015) T04001.
\bibitem{Werthm_14} D. Werthm\"uller  {\it et al.}, Phys. Rev. C                               {\bf 90}  (2014) 015205.
\end{thebibliography}
\end{document}